\documentclass[12pt]{article}
\usepackage{harvard}

\usepackage{setspace}

\usepackage{psfrag}                  
\usepackage{url}                        
\usepackage{times}                    
\usepackage{makeidx}               
\usepackage{graphicx}              
\usepackage{epstopdf}
\usepackage{multicol}                
\usepackage[bottom]{footmisc} 
\usepackage{fancyhdr}
\RequirePackage{subfigure}
\RequirePackage{rotating}
\RequirePackage{color}
\RequirePackage{makeidx}

\newcommand \be{\begin{equation}}
\newcommand \bea{\begin{eqnarray}}
\newcommand \ee{\end{equation}}
\newcommand \eea{\end{eqnarray}}

\renewcommand{\epsilon}{\varepsilon}

\makeindex

\begin{document}
\title{Look-Ahead Benchmark Bias\\ in Portfolio Performance Evaluation\thanks{We are grateful
to Y. Malevergne for useful discussions.}}
\author{
Gilles Daniel$^{1}$, Didier Sornette$^{1,2,\dag}$  and Peter Wohrmann$^3$\\
$^1$ ETH Z\"urich, Department of Management, Technology and Economics\\
Kreuzplatz 5, CH-8032 Z\"urich, Switzerland\\
$^2$ Swiss Finance Institute, c/o University of Geneva\\
40 blvd. Du Pont d'Arve, CH 1211 Geneva 4, Switzerland\\
$^3$ Swiss Banking Institute, University of Z\"urich\\
gdaniel@ethz.ch, dsornette@ethz.ch and peterw1@stanford.edu\\
$^\dag$ Contact author: Prof. Sornette, tel: +41 (0) 44 63 28917    ~  fax: +41 (0) 44 63 21914\\
www.er.ethz.ch
\\
}
\maketitle

\textbf{Abstract:} Performance of investment managers are evaluated
in comparison with benchmarks, such as financial indices. Due to the operational constraint that most
professional databases do not track the change of constitution of benchmark
portfolios, standard tests of performance suffer from the ``look-ahead benchmark bias,''
when they use the assets
constituting the benchmarks of reference at the end of the testing period, rather than at the
beginning of the period. Here, we report that the ``look-ahead benchmark bias''
can exhibit a surprisingly large amplitude for portfolios of common stocks (up to $8\%$ annum
for the S\&P500 taken as the benchmark) -- while most studies
have emphasized related survival biases in performance of mutual and hedge funds
for which the biases can be expected to be even larger. We use the CRSP database from 1926 to 2006 and
analyze the running top 500 US capitalizations to demonstrate that this bias can account for a gross overestimation of performance metrics such as the Sharpe ratio as well as an underestimation of risk, as measured for instance by peak-to-valley drawdowns. We demonstrate the presence of a significant
bias in the estimation of the survival and look-ahead biases studied in the literature. A general methodology to test the properties of investment strategies is advanced in terms of random strategies
with similar investment constraints.



\newpage

\section{Introduction}

Market professionals and financial economists alike strive to estimate the performance of mutual funds, of hedge-funds and more generally of any financial investment, and to quantify the return/risk characteristics of investment strategies. Having selected the funds and/or strategies of interest,
a time-honored approach consists in quantifying their past performance over some time period. A large literature has followed this route, motivated by the eternal question of whether some managers/strategies systematically outperform others, with its implications for market efficiency and investment opportunities.

Backtesting investment performance may appear straightforward and natural at first sight. However, a significant literature has unearthed, studied and tried to correct for ex-post conditioning biases, which include the survival bias, the look-ahead bias and data-snooping, which continue to pollute even the most careful assessments.
Here, we present a dramatic illustration of a variant of the look-ahead bias, that we refer to as
the ``look-ahead benchmark bias,'' which  surprised us by the large amplitude of the overestimation
of expected returns of up to $8\%$ per annum. This overestimation is comparable to the
largest amplitudes
of the survival biases and look-ahead biases found for mutual funds or hedge-funds.
We demonstrate the generic nature of the ``look-ahead benchmark bias'' by studying the
performance of portfolios investing solely in regular stocks using very simple strategies,
such as buy-and-hold, Markovitz optimization or random stock picking.

The look-ahead benchmark bias that we document is strongly related to the look-ahead bias
proper and to the survival bias, but has no particular relation to data-snooping \cite{Lo_etal90,White00,Sullivanetal99}, which we therefore do not discuss further.

The standard survivorship bias refers to the fact that
many estimates of persistence in investment performance are based on data sets that only contain funds that are in existence at the end of the sample period;
see, e.g. \cite{BrownGoetzmannIbbotsonRoss1992,GrinblattTitman92}. The corresponding
survivorship bias is caused by the fact that poor performing funds are less likely to be observed in
data sets that only contain the surviving funds, because the survival probabilities
depend on past performance. Perhaps less appreciated is the fact that stocks themselves
have also a large exit rate and hence also suffer from the survival  bias. For instance, \citeasnoun{Knaup05} examines the business survival characteristics of all establishments that started in the United States in the late 1990s when the boom of much of that decade was not yet showing signs of weakness, and finds that, if 85\% of firms survive more than one year, only 45\% survive more than four years. \citeasnoun{Bartelsmanetal03} confirm that a large number of firms enter and exit most markets every year in a group of ten OECD countries:
data covering the first part of the 1990s show the firm turnover rate (entry
plus exit rates) to be between 15 and 20 per cent in the business sector of most countries, i.e. a fifth of
firms are either recent entrants or will close down within the year. And this phenomenon
of firm exits is not confined to small firms. Indeed, in the exhaustive CRSP database of about 26'900 listed US firms, covering the period from Jan. 1926 to Dec. 2006 (Center for Research in Security Prices, \url{http://www.crsp.com/}), we find that on average 25\% of names disappeared after 3.3 years, 75\% of names disappeared after 14 years and 95\% of names disappeared after 34 years.

The standard look-ahead bias refers to the use of information in a simulation that would not be available during the time period being simulated, usually resulting in an upward shift of the results. An example is the false assumption that earnings data become available immediately at the end of a financial period.
Another example is observed in performance persistence studies, in which it is common
to form portfolios of funds/stocks based upon a ranking performed at the end of a first period,
together with the implicit or explicit condition that the funds/stocks are still in the
selected ranks at the end of the second testing period. In other words, funds/stocks that are considered for evaluation are those which survive a minimum period of time after a ranking period \cite{BrownGoetzmannIbbotsonRoss1992}. This bias is not remedied even if a survivorship free data-base is used, because it reflects additional constraints on ranking.

More generally,  the fact that a data set is survivorship free does not imply that standard methods of analysis do not suffer from ex-post conditioning biases, which in one way or another may use (often implicit or hidden) present information which would not have been available in a real-time situation.

Previous works have investigated both survivorship and look-ahead biases.
\citeasnoun{BrownGoetzmannIbbotsonRoss1992} have shown that survivorship
in mutual funds can introduce a bias strong enough to account for the strength of the
evidence favoring return predictability previously reported.
\citeasnoun{CarpenterLynch1999} find, among other results, that
look-ahead biased methodologies (which require funds to survive a minimum
period of time after a ranking period)  materially bias statistics.
\citeasnoun{ter Horst_etal01} introduce a
weighting procedure based upon probit regressions which models
how survival probabilities depend upon historical returns, fund age and
aggregate economy-wide shocks, and which provides
look-ahead bias-corrected estimates of mutual fund performances.
\citeasnoun{Baquero_etal05} apply the methodology of \citeasnoun{ter Horst_etal01}
to hedge-fund performance, which requires a well-specified model that
explains survival of hedge funds and how it depends upon
historical performance. \citeasnoun{ter Horst07} extend the  look-ahead bias correction
method of \citeasnoun{Baquero_etal05} to hedge-funds  by correcting separately for
additional self-selection biases that plague hedge-fund databases (underperformers do not wish to make their performance known while  funds that performed well have less incentive
to report to data vendors to attract potential investors).

The major part of the literature is devoted to assess the
look-ahead bias on actively managed investment funds. Here, we are
studying how the back-testing of investment strategies on biased
stock price databases is effected.
We add to the literature by focusing on the look-ahead bias that appears
when the assets used to test the portfolio performance are selected
on the basis of their relationship with the benchmark to which the performance is compared.
In the next section 2, we provide a specific straightforward implementation using the
S\&P500 index as the benchmark over the period from January 2001 to December 2006.
Section 3 is devoted to a more systematic illustration of the look-ahead benchmark bias over different periods, from 1926 to 2007. The substantial difference in the performances of up to
8$\%$ between portfolios with and without look-ahead bias provide an indication for the bias in the performance
of the back-test of an active investment strategy as it is
commonly carried out.
Sections 2 and 3 document that passive strategies are higher in after cleaning the
database with respect to the look-ahead bias.
Under quite general assumptions, we give in Section 4 an analytical
prediction for the look-ahead bias happening to the mean-variance
investment rule which might be applied in a mutual fund. In
particular, we discuss, under what conditions the naive
diversification would be favorable. The same methodology can be
applied to give decision support to the hedge fund manager whether
she should equally allocate money to different alternative
investment strategies. Section 5 extends on the empirical evidence presented
in sections 2 and 3 by using random strategies. Random strategies
are proposed as a simple and efficient test of the value
added by a given strategy, which take into account all possible biases, 
including those too difficult to address or that are even unknown to the analyst.

\section{A first recent illustration over the period from January 2001 to December 2006 using the S\&P500 as the benchmark}

\begin{figure}[ht]
\center
\includegraphics[width=12cm]{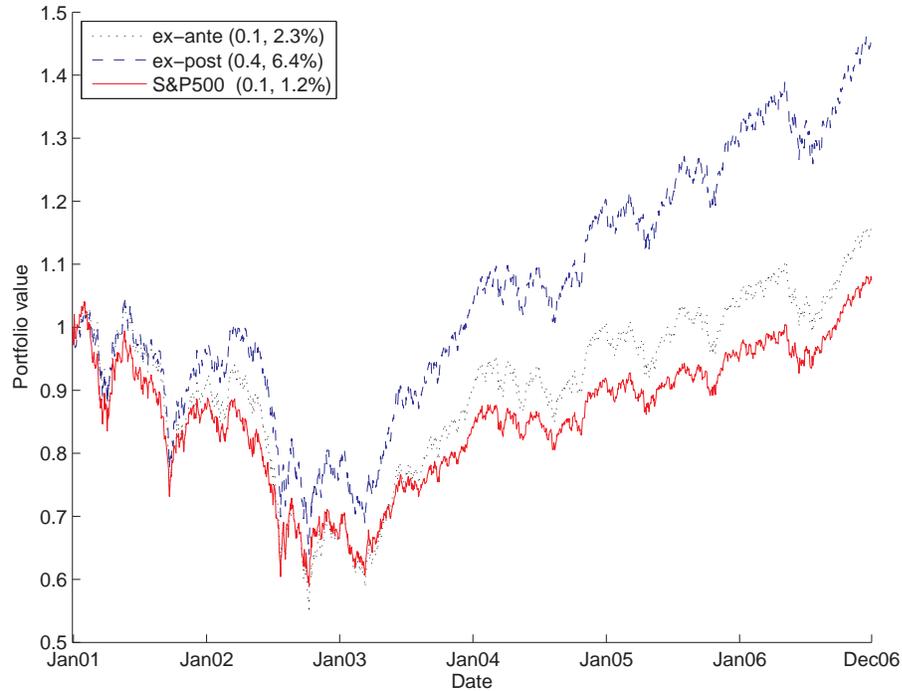}
\caption{Evolution, from January 2001 to December 2006, of \$1 invested in two equally weighted buy-and-hold portfolios made of the $500$ constituents of the S\&P500 index, respectively as in 1st January 2001 (ex-ante portfolio) and 29th December 2006 (ex-post portfolio). For reference, we also plot the historical value of the actual S\&P500 index, normalised to $1$ on 1st January 2001. The performance of these three portfolios are reported in the upper left panel, with their annualized Sharpe ratios (using a zero risk-free interest rate) and their continuously compounded average annual returns.}
\label{fig:Bias}
\end{figure}

Let us consider a manager who wants to back-test a given trading strategy on past data, namely on a pool of stocks such as the constituents of the S\&P500 index on a given period, say January 2001 to December 2006. To do so, the natural approach would be the following:
\begin{enumerate}
\item Obtain the list of constituents of the S\&P500 index at the present time (end of December 2006);
\item For each name (stock), retrieve the closing price time series for the given period from January 2001 to December 2006;
\item Backtest the strategy on that data set, for instance by comparing it with the S\&P500 benchmark.
\end{enumerate}
However, doing so introduces a formidable bias, and can easily lead to erroneous
conclusions. Figure \ref{fig:Bias} dramatically illustrates the effect by comparing
the performance of two investments.

In the first investment, which is subject to the look-ahead bias, we build \$1 of an equally weighted portfolio invested in the 500 stocks constituting the S\&P500 index {\it at the end of the period} (29 December 2006), and hold it from 1st January 2001 until 29th December 2006.
Meanwhile, the second investment simply consists in buying \$1
of an equally weighted portfolio invested in the 500 stocks constituting the S\&P500 index {\it at the
beginning of the period} (1st January 2001), and in holding it from 1st January 2001 until 29th December 2006. Both investments are buy-and-hold strategies and should have yielded similar performances if the
constitution of the S\&P500 index had not changed over this period.\footnote{Since the S\&P500 index is *not* equally weighted, we should expect a slight discrepancy between the evolution of portfolio 2 and the actual index.}
However, the list of constituents of the S\&P500 index is updated, usually on a monthly basis, in order to
include only the largest capitalizations of the US stock market at the current time. Consequently, this list of constituents cannot, almost per force, contain a stock that for instance crashed in the recent past. Indeed, in such a case, the stock has a large probability to be passed in terms of capitalization by another stock of the same industry branch and thus leave the index and be replaced by that other stock. The only difference in the two portfolios is that the first investment uses a look-ahead information, namely it knows on 1st January 2001 what will be the list of constituents of the S\&P500 index at the end of the period (29th December 2006).
This apparently innocuous look-ahead bias leads to a huge difference in performance, as can be seen in Figure \ref{fig:Bias} and from simple statistics: the first (respectively second) investment has an annual average compounded return of $6.4\%$ (resp. $2.3\%$) and  a Sharpe ratio (non-adjusted for risk-free rate) of $0.4$ (resp. $0.1$). The first investment has significant better return but, what is even more important, it exhibits larger risk-adjusted returns.

 For reference, we also plot the historical value of the actual S\&P500 index, normalised to $1$ on 1st January 2001. Note that its performance is slightly worse than that of the second investment discussed above. This could be due to the different weighting and also to the effect reported by \citeasnoun{Cai_Houge07}\footnote{The look-ahead benchmark bias documented here is related to the work of \citeasnoun{Cai_Houge07} who study how additions and deletions  affect benchmark performance. Studying changes to the small-cap Russell $2000$ index from 1979-2004, \citeasnoun{Cai_Houge07} find that a buy-and-hold portfolio significantly outperforms the annually rebalanced index by an average of 2.2\% over one year and by 17.3\% over five years. These excess returns result from strong positive momentum of index deletions and poor long-run returns of new issue additions.}.

Many managers would have been happier to report Sharpe ratios in the range obtained for the first investment, especially over this turbulent time period. Investment strategies exhibiting this kind of performance would fuel interpretations that this is evidence of a departure from the efficient market hypothesis and/or of the existence of arbitrage opportunities. On the other hand, other pundits would observe that this look-ahead bias is really obvious, so that no one would fall into such a trivial trap. This quite reasonable assessment actually collides with one simple but often overlooked operational limitation
of back-tests\footnote{Since the benchmark is observed continuously, real-time assessment of performance does not of course suffer from this problem. We only refer to back-testing which uses
a recorded time series of the benchmark and present knowledge of its constituents.}: the change of constitution of financial indices are not recorded in most standard professional databases, such as Bloomberg, Reuters, Datastream  or Yahoo! Finance.
As the standard goal for investment managers is at least to emulate or better to beat some index of reference, back-tests on comparative investments should use a defined set of assets on which to invest, which is defined {\it at the beginning of the period}. However, because the list of constituents of the indices
is unexpectedly challenging to retrieve\footnote{Standard \& Poor's themselves provide the list of constituents of the S\&P500 index only since January 2000, while scripting Reuters, Bloomberg and Datastream returned only incomplete results. In fact, it appears that both the CRSP and Compustat databases are necessary to retrieve the list of constituents of the S\&P500 index at any given point in time, and these databases are usually not accessible to practitioners.}, it is common practice to use the set of assets constituting the benchmarks of reference at the present time, rather than at the
beginning of the period. Then, necessarily, the kind of look-ahead bias that we report here will automatically pollute the conclusions, with sometimes dramatic consequences, as illustrated in Figure \ref{fig:Bias}. We refer to this as the ``look-ahead benchmark bias.''

A part of the over-performance of the ex-post portfolio over the S\&P500 index
can be attributed to the fact that the former is equally-weighted while the later
is value-weighted. But this does not explain away the look-ahead effect
as shown by the large difference between the equally-weighted ex-post
and ex-ante portfolios. For instance, consider the DJIA. While the
mean return of the price-weighted DJIA index from January 2001 to September 2007 was on slightly below
($3.2\%$ p.a) that of the  price-weighted ex-post portfolio ($3.8\%$ p.a.),
the difference is much larger for the period from February 1973 to September 2007:
$5.7\%$ p.a. versus $7.8\%$ p.a.

\section{The extent of the Look-Ahead Benchmark Bias}

We use the data provided by the  CRSP database, from which we extract the close price (daily), split factor (daily) and number of outstanding shares (monthly) for all US stocks from January 1926 to December 2006. This represents a total of 26'892 different stocks.

\begin{figure}[ht]
\center
\subfigure[biased, ex-post pickup]{\includegraphics[width=10cm]{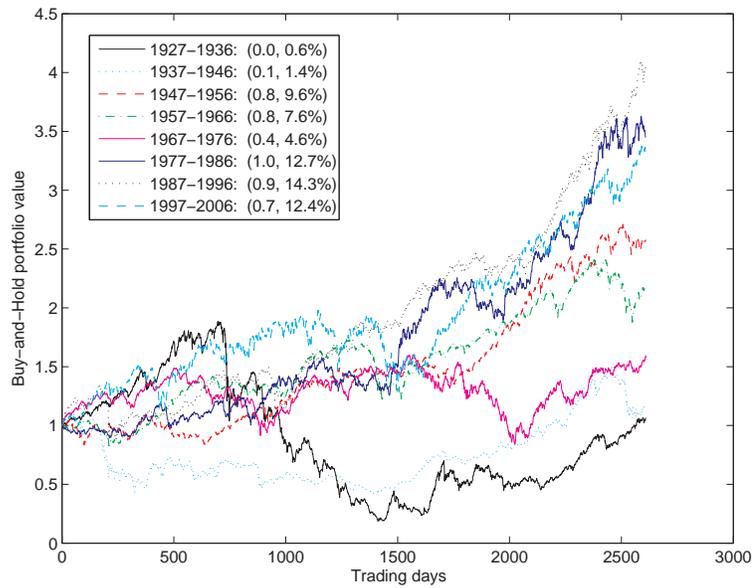}}
\subfigure[ex-ante pickup]{\includegraphics[width=10cm]{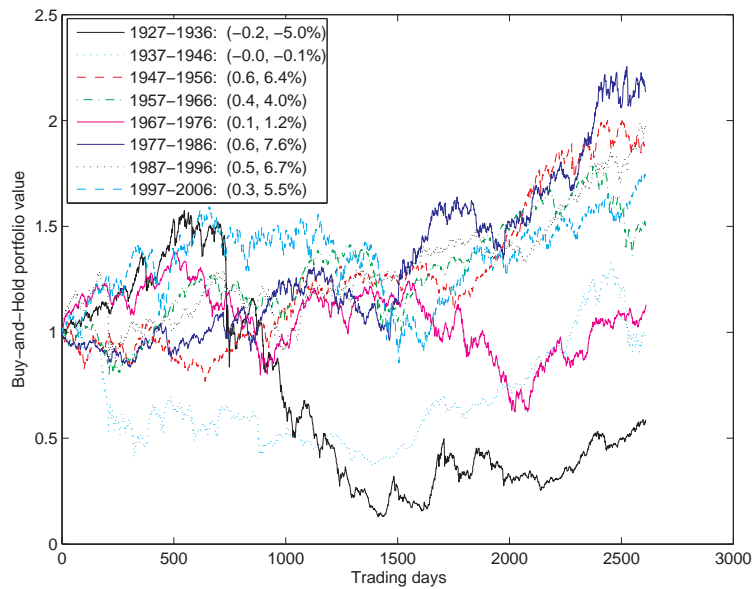}}
\caption{For each epoch of $10$ years, we plot the evolution of portfolio $1$ (upper panel)
which invests \$1  equally weighted on the $500$ largest stock capitalizations, as determined at the end
of the ten-year period and the evolution of portfolio $2$ (lower panel)
which invests \$1  equally weighted on the $500$ largest stock capitalizations, as determined at the start
of the ten-year period. Note the different ranges of the vertical scales in the two panels.
The inserts give the Sharpe ratio (with zero risk-free rate) and continuously-compounded average annual return. The discrepancy between these two figures helps us visualise the extent of the survival bias for the $500$ largest capitalizations throughout time.}
\label{fig:500largest}
\end{figure}

We decompose the time interval from January 1926 to December 2006 into $8$ periods of $10$ years
each. For each period, we monitor the value of two portfolios. At the beginning of each ten-year period, the first ex-post (resp. second ex-ante) portfolio invests \$1 equally weighted on the $500$ largest stock capitalizations, as determined at the end (resp. start) of the ten-year period. The first ex-post portfolio has by definition the look-ahead benchmark bias, while the second ex-ante portfolio is exempt from it and could have been implemented in real time.
Figure \ref{fig:500largest} plots the time evolution of the value of the two portfolios. The inserted panels show that the Sharpe ratio and continuously-compounded average annual returns are much larger for portfolio $1$ compared with portfolio $2$.

\begin{figure}[ht]
\center
\includegraphics[width=12cm]{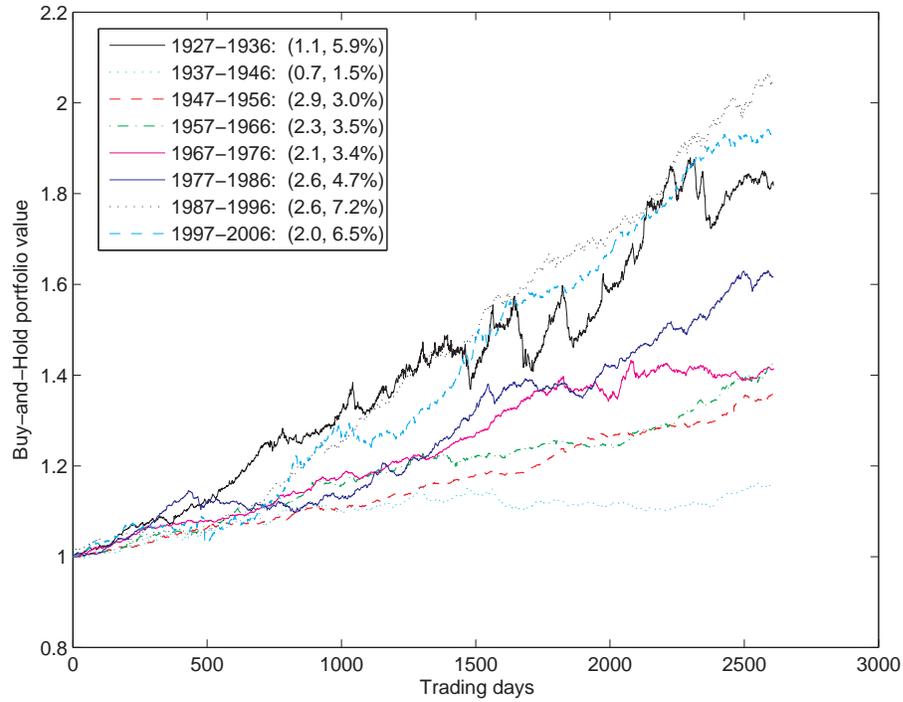}
\caption{Time evolution of the value of an investment consisting in being long \$1
in a portfolio $1$ equally weighted on the 500 largest stock capitalizations at the end of each period
and short \$1 in a portfolio $2$ equally weighted on the 500 largest stock capitalization at the end of each period, with compounding the returns. The inset shows the Sharpe ratios (with zero risk-free interest) and compounded annual return for the $8$ periods.}
\label{fig:ratio-500largest}
\end{figure}

Figure \ref{fig:ratio-500largest} shows the time evolution of the value of an investment consisting of being long \$1 in the portfolio $1$ (ex-post) and short \$1 in portfolio $2$ (ex-ante). In other words, it shows  the ratio of the value of the ex-post portfolio $1$ divided by the value of the ex-ante portfolio $2$, for each of the $8$ periods. By construction, this hedged long-short portfolio can be implemented only ex-post when backtesting, and not in real-time. Its performance is consistently good \footnote{Only the 1937-1946 period exhibits a rather smaller gain, albeit still positive with a significant reduction of risks.} over the $8$ periods from 1926 to 2006, with a huge reduction of risks and better returns than the unbiased, ex-ante portfolio $2$, thus demonstrating the significance of the look-ahead bias. This result is robust with respect to the number of stocks selected in the two portfolios.

Table \ref{table:returntest} tests for different means in the ex-ante and ex-post portfolios. In the
first two decades the means cannot be distinguished while in the
recent decades the means differ significantly.

\section{Theoretical estimation of the bias in estimations of the look-ahead and survival biases}

We now show how one can analytically determine the estimation bias with
regard to the true value of the bias, using the fact that
the sample mean of an (equally weighted or Markowitz) portfolio of
assets with normally distributed returns has a Wishart
distribution. We find in particular that
the level of the entries in the covariance matrix of the asset returns has an impact on the
amount of the bias. This is relevant because a database
with survival bias has a covariance matrix with smaller covariance terms,
which tends to enhance the difference between the true and the biased dataset.
These calculations can also be used for the survival bias, as discussed at the end of the
section.

In the following, we calculate analytically the estimation error of the Sharpe Ratio based on data with bias versus the Sharpe Ratio based on data without bias. The derived expression depends on the true expected returns and covariance matrices of both data sets.  Let us consider an economy
characterized by a vector process of $N$ asset excess returns $\{R_t, t=1,..., T\}$, which
are normally distributed. Let $\mu$ be the vector of mean excess return, $\Sigma$ the
covariance matrix of the excess returns and $\omega$ the vector of portfolio weights.

Markowitz' optimization program consists in finding $\omega$ maximizing the
following risk-adjusted excess return
\be
U(\omega) = \omega' \mu  - {\gamma \over 2} \omega' \Sigma \omega~,
\ee
where $\gamma$ is the risk aversion coefficient. The optimal weights are
\be
\omega^* = {1 \over \gamma} \Sigma^{-1} \mu~,
\ee
with
\be
U(\omega^*) =  {1 \over 2 \gamma} \omega' \Sigma^{-1} \omega = {1 \over 2 \gamma} S_*^2~,
\ee
where $S_*$ is the Sharpe ratio given by
\be
S_*^2 = \omega' \Sigma^{-1} \omega~.
\ee
In contrast, the naive (equally weighted) diversification $\omega_{\rm EW} = c {\bf 1}_N$, where $c \in {\cal R}$
gives the following risk-adjusted excess return
\be
U(\omega_{\rm EW}) =  {1 \over 2 \gamma} S_{\rm EW}^2~,
\ee
where the Sharpe ratio $S_{\rm EW}$ of the equally weighted portfolio is given by
\be
S_{\rm EW} = {({\bf 1}' \mu)^2 \over {\bf 1}_N' \Sigma {\bf 1}_N}~.
\ee

 Sample estimations of the mean excess return, covariance matrix and optimal Markowitz' weights
 are
 \be
 {\hat \mu} = {1 \over T} \Sigma_{t=1}^T R_t~,~~~{\hat \Sigma} = {1 \over T}
 \Sigma_{t=1}^T (R_t -  {\hat \mu})(R_t -  {\hat \mu})'~,~~~{\hat \omega} = {1\over \gamma}
 {\hat \Sigma}^{-1} {\hat \mu}~.
 \ee
Then, the sample excess returns $ {\hat \mu}$ are distributed according to a multivariate
normal distribution,
\be
{\hat \mu} \sim {\cal N}\left(\mu, {\Sigma \over T}\right)~,
\ee
and the sample covariance matrix is distributed according to
\be
T {\hat \Sigma} \sim {\cal W}_N\left(\Sigma, T-1\right)~,
\ee
where ${\cal W}_N$ is the Wishart distribution.

Let the index ``1'' refer as in the above empirical tests to data with the look-ahead bias,
and the index ``2'' to data without the bias. The true bias is $U(\omega_1^*) - U(\omega_2^*)$,
while one only has access to the estimated bias $U({\hat \omega}_1) -  U({\hat \omega}_2)$.
To access how much the bias can be under- or over-estimated, the relevant
measure is
\be
\Delta \equiv \left[U(\omega_1^*) - U(\omega_2^*)\right]
- \left[U({\hat \omega}_1) - U({\hat \omega}_2)\right]~,
\ee
which can be expressed explicitly as
\be
\Delta = {1 \over \gamma}  (1-k) \left[S_{*,1}^2  - S_{*,2}^2\right]
=  {1 \over \gamma}  (1-k) \left(  \omega_1' \Sigma_1^{-1} \omega_1
- \omega_2' \Sigma_2^{-1} \omega_2 \right]~,
\ee
where
\be
k = {T \over T-N-2} \left(2 - {T(T-2) \over (T-N-1)(T-N-4)}\right) <1~.
\ee
Suppose that the biased data is such that ${\bf 1}' \mu_1 > {\bf 1}' \mu_2$ and
${\bf 1}_N' \Sigma_1 {\bf 1}_N < {\bf 1}_N' \Sigma_2 {\bf 1}_N$, which is indeed
usually the case as reported above (large returns and smaller risks
for the look-ahead biased data). Then, $\Delta >0$, i.e.,
the true bias is larger than estimated from the data. If the sample size $T$ is not too large compared with the number $N$ of assets, then the effect of the bias on the Sharpe Ratio is generally underestimated.
This under-estimation is also found for the equi-weighted portfolio but its magnitude is larger for the Markowitz rule.

Other effects occur. For instance, consider portfolios made of a few tens of assets with the same expected returns and same variances, and with zero covariances. Then, the expected Sharpe ratio is slightly better for na\"ive
diversification as compared to the sample-based Markowitz-optimal portfolio. Now, assume you have measured the expected returns and their covariance matrix
based on data with the survival bias, where the mean and variances are
higher than in bias-free data. Then, the expected measure of the Sharpe
ratio gets higher for the sample-based Markowitz-optimal portfolio compared
to the na•ve diversification, that is, the order flips over.
In the literature on the survival bias, the methodology is often to compare the performance measures estimated by sample versions of expected returns and their covariance matrix based on both a clean and a biased data base. As our calculations above show (and which apply straightforwardly
also the survival bias), the assessment of the impact of the survivorship bias on the performance figures is biased itself. Under mild assumptions, one can conclude that the bias is worse than one thinks  it is when reading the literature, which includes e.g. \cite{BrownGoetzmannIbbotsonRoss1992,Brownetal95,Brownetal97,Brownetal99,Carhartetal02,CarpenterLynch1999,Eltonetal96}.

\section{Constrained Random Portfolios and proposed testing methodology}

\subsection{Robustness of the look-ahead benchmark bias using random strategies}

Let us come back to the period from 1st January 2001 to 29th December 2006,
shown in Figure \ref{fig:Bias} in order to test further the amplitude and
robustness of the impact of the look-ahead benchmark bias.
Investing on the 500 constituents of the S\&P500 index at the end
of the testing period (29th December 2006) amounts to bias the stock selection
towards good performers. We illustrate that, as a consequence, non-informative
random strategies exhibit very good to extremely good performance.
We generate 10 portfolios, each of them implementing a random strategy. A random strategy opens only long positions (we buy first, and sell later) on a subset of the 500 stocks, with an average leverage of $0.8$ and an average duration per deal of $9$ days -- which are common values. Given these constraints, the choice of the selected stock and the timing are random at each time step. We do not specify further the algorithm as all possible specific implementations give similar results.

\begin{figure}[ht]
\center
\includegraphics[width=12cm]{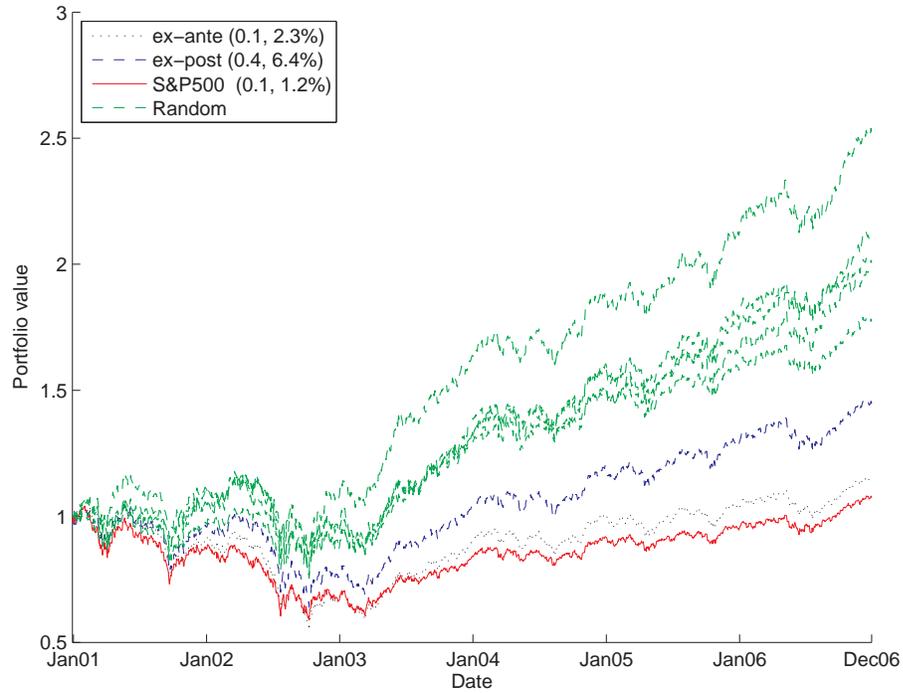}
\caption{Time  evolution of the value of \$1 invested
in each of five random portfolios (as described in the text)  on 1st January 2001. The random portfolios (solid green lines) strongly outperform the S\&P500 index (red dashed
line), and simply diffuse around the look-ahead index  (black dots) with an upward asymmetry.
 }
\label{fig:RandomPortfolios}
\end{figure}

Figure \ref{fig:RandomPortfolios} plots the time evolution of the value of \$1 invested
in each of these ten random portfolios on 1st January 2001. These random portfolios provide
an average compounded annual return of $(9.1 \pm 4) \%$ with an excellent Sharpe
ratio (with zero risk-free interest) of $2 \pm 0.8$.
The random portfolios (solid green lines) strongly outperform the S\&P500 index (red dashed
line), and simply diffuse around the look-ahead index  (black dots) with an upward asymmetry.
Similar results are obtained with other parameters of the random strategies.

\subsection{Using random strategies on the same biased data as a general testing procedure}

Because in practice one may never be in a position to completely exclude the presence of a look-ahead bias, we propose the use of constrained random portfolios on the same database as benchmarks
against which to test the value of proposed investment strategies, so as to
to assess the probability that the performance of a given strategy can be attributed to chance.
Because the same look-ahead
biases will impact both the random portfolios and the proposed strategies, one should thus
be able to detect the presence of anomalously large gains that could result from a large
amplitude of the look-ahead bias and quantify the real value, if any, of the proposed
strategy over the random portfolios. This added value can be used as a useful metric
of the performance of the proposed strategy. In order for this methodology to work, the
constrained random portfolios should imitate as closely as possible the properties of the trading strategy about to be tested, such as its leverage, mean invested time and the turn-over.

\section{Concluding remarks}

We have reported a surprisingly large ``look-ahead benchmark bias'', which results from
an information on the future ranking of stocks due to their belonging to a benchmark
index at the end of the testing period. We have argued that this look-ahead benchmark
bias is probably often present due to (i) the need for strategies or investments to prove
themselves against benchmark indices and (ii)
the non-availability of the changes of
composition of benchmark indices over the testing period.
More generally, it is difficult if not impossible to completely
exclude any look-ahead bias in simulations of the performance of investment strategies.
As we reviewed in the introduction, one way to address these biases requires to
first recognize their existence, and then to model how survival probabilities depend on
historical returns, fund age and aggregate economy-wide shocks.

But, one can never be 100\% certain that all biases have been removed.
In certain scientific fields concerned with forecasting, such as in earthquake prediction for instance,
the community has recently evolved to the recognition that only real-time procedures can avoid
such biases and test  the validity of models \cite{Jordan06,Schorlemmer07}.
Actually, real-time testing is a standard of the financial industry, since cautious investors only
invest in funds that have developed a proven track record established over several years. But as shown
in the academic literature, success does not equate to skill and may
not be predictive of future performance, as luck and survival bias can also loom
large \cite{Barrasetal07}. There is a large and growing literature on
how to test for data snooping and fund performance (see for instance
\cite{Lo_etal90,Romano_Wolf05,Wolf06}). The problem is more generally related to the
larger issue of validating models (e.g., \cite{Sornette_etal07} and references therein).

Practitioners should be careful to test for the presence of such a look-ahead bias in their dataset prior to backtesting their trading strategy. We have proposed a simple and practical way to do so, which
consists in using random strategies, having as much of the characteristics of the strategy to be
tested as is possible.


\clearpage

\begin{table}[ht]
\caption{Two-sample-t-test for indentical means (with unknown but common standard deviation)
for the ex-post and ex-ante portfolios 1 and 2. The test uses the real one year interest rates as reported by R. Shiller. }
\vskip 0.2cm
\centering 
\begin{tabular}{c c c } 
\hline\hline 
Sub-sample & Test statistic & Significance level  \\ [0.5ex] 
\hline 
1927-1936  &    0.7201  &           0.4807  \\
1937-1946  &    0.3708  &           0.7151\\
1947-1956  &    1.4125  &         0.1749\\
1957-1966  &    1.9151  &           0.0715\\
1967-1976  &    3.2103  &         0.0049\\
1977-1986  &    2.3461  &           0.0306\\
1987-1996  &    2.6059  &           0.0179\\
1997-2006  &    1.9879  &           0.0622\\ [1ex] 
\hline 
\end{tabular}
\label{table:returntest} 
\end{table}

\clearpage

\clearpage

\clearpage

\clearpage

\end{document}